# Identifying Semantic Similarity for UX Items from Established Questionnaires Using ChatGPT-4

Stefan Graser
*Center for Advanced E-Business Studies*
*RheinMain University of Applied Sciences*
Wiesbaden, Germany
ORCID: 0000-0002-5221-2959
stefan.graser@hs-rm.de

Martin Schrepp
*SAP SE*
Walldorf, Germany
ORCID: 0000-0001-7855-2524
martin.schrepp@sap.com

Stephan Böhm
*Center for Advanced E-Business Studies*
*RheinMain University of Applied Sciences*
Wiesbaden, Germany
ORCID: 0000-0003-3580-1038
stephan.boehm@hs-rm.de

*Abstract*—Questionnaires are a widely used tool for measuring the user experience (UX) of products. There exists a huge number of such questionnaires that contain different items (questions) and scales representing distinct aspects of UX, such as efficiency, learnability, fun of use, or aesthetics. These items and scales are not independent; they often have semantic overlap. However, due to the large number of available items and scales in the UX field, analyzing and understanding these semantic dependencies can be challenging. Large language models (LLM) are powerful tools to categorize texts, including UX items. We explore how ChatGPT-4 can be utilized to analyze the semantic structure of sets of UX items. This paper investigates three different use cases. In the first investigation, ChatGPT-4 is used to generate a semantic classification of UX items extracted from 40 UX questionnaires. The results demonstrate that ChatGPT-4 can effectively classify items into meaningful topics. The second investigation demonstrates ChatGPT-4's ability to filter items related to a predefined UX concept from a pool of UX items. In the third investigation, a second set of more abstract items is used to describe another classification task. The outcome of this investigation helps to determine semantic similarities between common UX concepts and enhances our understanding of the concept of UX. Overall, it is considered useful to apply GenAI in UX research.

## I. INTRODUCTION

Questionnaires designed to measure the user experience (UX) of products contain items that allow users to judge how effectively the product supports important aspects of user interaction and expectations. It is important to note that items in such questionnaires are semantically not independent. With the recent advances in large language models, such as ChatGPT-4, we now have the opportunity to explore the semantic similarities of UX items in a more efficient and structured manner. We enhance in this paper the first approaches [1] to use ChatGPT-4 for structuring UX items with new data and methods.

User Experience (UX) is a holistic concept in Human-Computer-Interaction (HCI) that refers to the subjective perception of users regarding the use and interaction with a product, service, or system [2]. Ensuring a good level of UX is important for the long-term success of products and services. Therefore, the perception of the users regarding UX must be investigated and measured to collect insights that can be used to enhance the UX [3]. Various methods, for example, usability tests or expert reviews, allow us to gain insights into the UX of a product. However, the most commonly used approach to measure UX is through standardized questionnaires gathering self-reported data from users [4]. These questionnaires can be applied in a cost-efficient, simple, and fast way [4][5].

UX is a complex concept that encompasses various aspects, including efficiency, learnability, enjoyment, aesthetic appeal, trust, and loyalty, among others. Since the number of questions that can be asked in a survey is limited, a single UX questionnaire can not cover all aspects comprehensively. This is why there are numerous UX questionnaires available, each optimized to address specific research questions through its items and scales. Each questionnaire measures only a subset of the potential UX aspects. Attempts to compare different UX questionnaires and to help practitioners select the most suitable one for their research questions are described in [6]–[8].

The existing UX questionnaires have been developed by various authors over a long period of time. As a result, it is not surprising that there is no consensus on the factors and items included in standardized UX questionnaires. Factors with different names may measure the same thing, while factors with the same name may measure different aspects [9]. Therefore, it is necessary to carefully examine the individual items of a scale in a questionnaire to gain a clear understanding of its meaning and potential overlap with other scales from the same or different questionnaires.

Thus, a semantic analysis of UX items from questionnaires can help to develop a deeper understanding of the meaning of UX scales. In this article, we investigate whether Generative AI, specifically ChatGPT-4, can be utilized for this purpose. We used ChatGPT-4 to analyze and compare items from existing UX questionnaires concerning their semantics. Based on this, similar items can be clustered, items representing a specific research question can be determined, and the semantic relation of commonly used UX concepts can be visualized. With this context in mind, we address the following research questions:

**RQ1:** *Is Generative AI able to generate a meaningful semantic classification of existing UX items?*

**RQ2:** *Is Generative AI able to filter items representing a predefined UX concept out of a pool of existing UX items?*

**RQ3:** *Can Generative AI help to uncover semantic similarities between common UX concepts and help to understand the concept of UX better?*

This article is structured as follows: Section II describes the theoretical foundation of our approach. Section III shows related work concerning the consolidation of UX factors and common ground in UX research. In addition, the research objectives are specified. Section IV illustrates the methodological approach of this study applying ChatGPT in UX research. Based on this, Sections V, VI, and VII show the three main investigations and the respective results answering the three research questions. A conclusion and outlook are given in Section VIII.

## II. THEORETICAL FOUNDATION

### A. Concept of UX

As mentioned in the introduction, UX is a multi-faceted concept that encompasses various aspects of product quality. It is important to distinguish between the traditional concept of usability, which is defined as "the extent to which a product can be used by specified users to achieve specified goals with effectiveness, efficiency, and satisfaction in a specified context of use" [2] and the modern concept of UX. Usability is focused on completing tasks and achieving goals with a product. UX, on the other hand, encompasses a broader spectrum of qualities that contribute to the subjective impression of a product. This includes, for example, aspects such as aesthetics or fun of use that are not directly connected to solving tasks with a system. In this sense, usability can be declared a subset of UX [8][10][11].

Hassenzahl established a distinction between pragmatic and hedonic UX qualities. Pragmatic qualities are task-related, while hedonic qualities are non-task-related [12]. However, this distinction poses some challenges. Firstly, whether some UX aspects are pragmatic or hedonic is not always clear. For example, content quality is obviously important for most web pages. If users search for specific information on a page, then high-quality content helps them find answers quickly, making it a pragmatic UX quality. On the other hand, if users stumble upon the page while browsing without a specific goal in mind, high-quality content becomes more of a hedonic quality. Secondly, pragmatic qualities adhere to a common concept as they are task-related, whereas hedonic qualities do not follow such a concept [13]; they simply encompass the remaining qualities that do not fit into the category of pragmatic qualities.

In [13], UX is conceptualized through a set of quality aspects. The basic concept is explained by defining that a "UX quality aspect describes the subjective impression of users towards a semantically clearly described aspect of product usage or product design". These UX quality aspects relate either to the external goals of the user (for example, to finish work-related tasks quickly and efficiently), to psychological needs (for example, fun of use or stimulation), sensory qualities (for example, the tactile experience when operating a device) or simply by the needs of the manufacturer to promote the product (for example, that the design looks novel and creative to attract the attention of potential customers) [13].

### B. Semantic and Empirical Similarity

Our goal is to uncover the semantic similarity between items in UX questionnaires. We understand semantic similarity as the degree of likeness or resemblance between the item texts based on their meaning. In this sense, semantic similarity goes beyond surface-level syntactic or structural similarity and takes into account the context, relationships, and associations between words or phrases to determine their level of similarity [14]–[16]. Different statistics-based methods in Natural Language Processing (NLP) to measure Semantic Textual Similarity are described in the research literature [14][17]–[24]. These methods can be divided into Matrix-Based Methods, Word Distance-Based Methods, and Sentence Embedding Based Methods [25].

Large Language Models, like GPT, use word embeddings (dense vector representations of words derived with the help of deep learning mechanisms applied to vast volumes of existing texts) to calculate semantic similarity. Therefore, they are a natural choice for analyzing the semantic similarity of UX items.

However, to fully understand also the limitations of such an approach, we need to take a closer look at the relation between semantic similarity and empirical similarity of items [26][27]. In survey research, the empirical correlation of items or scales is typically used to describe how closely related they are and if they measure similar constructs. However, we may observe in studies that items with a small semantic similarity, as estimated by an LLM, show quite substantial correlations.

A well-known example is the observation that beautiful products are perceived as usable [28][29]. Thus, a substantial correlation often exists between items that measure beauty and classical usability items. Thus, visual aesthetics influence the perception of classical UX aspects like efficiency, learnability, or controllability. A similar effect exists also in the opposite direction, i.e., the perception of usability influences the perception of beauty [30][31].

Several psychological mechanisms (which, in fact, may all contribute to the effect) have been proposed to explain these unexpected empirical dependencies, for example, the general impression model [32], evaluative consistency [33], or mediator effects [34]. Another explanation is that aesthetics and usability share common aspects. It is well-known that balance, symmetry, and order [35] or alignment [36] influence the aesthetic impression. However, a user interface that looks clean, ordered, and properly aligned is also easy to scan and navigate. Users can find information faster and orient themselves more easily on such an interface. Hence, balance, symmetry, and order will also benefit efficiency or learnability [27].

Items with a high semantic similarity address similar UX aspects, and participants in a survey should give highly similar answers to such items. Therefore, items with a high semantic similarity will also show empirically high correlations. But, the converse is not always true. There may be items with low semantic similarity but substantial empirical correlations due to the aforementioned effects. Thus, we should not expect to reconstruct scales of established questionnaires by a purely semantical analysis of the items. Typically such scales are developed by an empirical process of item reduction, mostly by main component analysis, and grouping items into scales based on their empirical correlations observed in larger studies.

*C. UX Questionnaires*

User experience refers to the subjective perceptions of users towards a product or system. Therefore, it is essential to gather feedback directly from users. Theoretical evaluations of UX based solely on system properties are not feasible. Since they are easy to set up and allow for the asking of many users with low effort, survey-based methods are currently the most frequently used method for quantitative UX evaluation. To ensure meaningful and comparable results, it is crucial to incorporate standardized UX questionnaires into these surveys. Additionally, collecting demographic information about participants, providing comment fields, or including specific questions can further enhance the evaluation process.

In recent decades, several standardized UX questionnaires have been developed. For instance, [9] provides a description of 40 common UX questionnaires, and an even longer list is presented in [8]. It is important to note that UX is a multifaceted concept, and no single questionnaire can cover all aspects of it. Thus, every questionnaire is based on specific UX quality aspects, which are represented as scales in the questionnaire. Each scale is represented by a number of items (questions) that correspond to the UX aspect being measured by the scale. The choice of the most suitable questionnaire for a given research question heavily depends on the specific UX quality aspects that are most relevant in that particular case. For example, when evaluating a product primarily used for professional work, classical usability aspects such as efficiency, learnability, and dependability are of high importance. Consequently, the questionnaire used for evaluation should include corresponding scales that measure these aspects. On the other hand, if the goal of the evaluation is to compare two versions of a product in terms of their visual design, a specialized questionnaire that focuses on this aspect, such as the VISAWI [37], is a better choice. Now, let's examine some examples of UX questionnaires and their item formats.

Díaz-Oreiro et al. [38] reported that the User Experience Questionnaire (UEQ) [39] is currently the most widely used questionnaire for UX evaluation. The UEQ developed by [39] is based on the distinction of UX aspects into pragmatic and hedonic scales [12][39]. The questionnaire consists of six scales:

- **Attractiveness**: Overall impression of the product. Do users like or dislike it?
- **Perspicuity**: Is it easy to get familiar with the product and to learn how to use it?
- **Efficiency**: Can users solve their tasks without unnecessary effort? Does it react fast?
- **Dependability**: Does the user feel in control of the interaction? Is it secure and predictable?
- **Stimulation**: Is it exciting and motivating to use the product? Is it fun to use?
- **Novelty**: Is the design of the product creative? Does it catch the interest of users?

The scales Perspicuity, Efficiency, and Dependability are pragmatic scales, Stimulation, and Novelty are hedonic scales, and Attractiveness is a pure valence scale (overall impression, which does not relate to concrete properties of the interaction between user and product) [39].

Each scale consists of four items. The items are semantic differentials with a 7-point Likert scale, i.e., each item consists of a pair of opposite terms that represent a UX dimension, for example, *inefficient - efficient*, *confusing - clear*, *not interesting - interesting* or *conventional - inventive*. Further details can be found online [40].

Many other questionnaires (especially the questionnaires that focus on usability, i.e., task-related UX quality) employ a different measurement concept. These questionnaires contain items that pertain to specific interface elements. For example, the Purdue Usability Testing Questionnaire [41] contains items like "Is the cursor placement consistent?" or "Does it provide visually distinctive data fields?". Other questionnaires use more abstract statements about the product to which the participants can express how much they agree or disagree on an answer scale, for example, "I found the system unnecessarily complex" (from the System Usability Scale [42]) or "The software provides me with enough information about which entries are permitted in a particular situation" or "Messages always appear in the same place" (from ISOMETRICS [43]). This type of item is more concrete but can only be applied to a certain type of product. In addition, there are several questionnaires that can be applied only for special application domains, for example, web pages, e-commerce, or games (for an overview of common questionnaires and item formulations, see [8]). The diverse formulation of items in UX questionnaires makes it challenging to categorize them based on their semantic meaning.

As previously mentioned, each UX questionnaire focuses on a specific subset of all possible UX quality aspects. Therefore, it is common practice to combine or utilize multiple questionnaires simultaneously in order to cover all relevant aspects required to answer a specific research question. However, due to the presence of different items and scales, participants may find it more challenging to complete the evaluation. Therefore, [44] developed the UEQ+, a modular framework. The framework is based on described factors with their respective items covering the construct UX as broadly as possible. Researchers can choose from a set of 27 UX quality aspects according to the respective product to evaluate and create an individualized UX questionnaire. Further information can be found online [44][45].

III. RESEARCH OBJECTIVE AND RELATED WORK

Due to the various UX questionnaires, many different factors and items exist. Hence, a lack of common ground in breaking down the concept of UX can be shown in the field of quantitative UX evaluation. Against this background, only a little research was done to consolidate general UX factors and, thus, find a common understanding. This results in a respective research gap. Only three records concerning a consolidation based on empirical similarity and two records in relation to semantic similarity can be found in the literature. In the following, we present the respective approaches.

Regarding a consolidation based on empirical similarity, [46] can be first listed. [46] analyzed existing questionnaires from the literature and consolidated the collected factors based on their definition. This resulted in a consolidated list of general UX factors [46]. The second approach by [6] was based on this. In this article, [6] also conducted the consolidation based on the definitions as well as experts. The latest approach was done by [13], resulting in a list of consolidated UX factors. In this context, the term UX quality aspect (See Section II-A) was introduced and can be considered equivalent to the term UX factor. The UX quality aspects based on [13] are shown in the following table (see Table I).

TABLE I: CONSOLIDATED UX FACTORS BASED ON [13].

| (#) | Factor | Descriptive Question |
|---|---|---|
| (1) | Perspicuity | Is it easy to get familiar with the product and to learn how to use it? |
| (2) | Efficiency | Can users solve their tasks without unnecessary effort? Does the product react fast? |
| (3) | Dependability | Does the user feel in control of the interaction? Does the product react predictably and consistently to user commands? |
| (4) | Usefulness | Does using the product bring advantages to the user? Does using the product save time and effort? |
| (5) | Intuitive Use | Can the product be used immediately without any training or help? |
| (6) | Adaptability | Can the product be adapted to personal preferences or personal working styles? |
| (7) | Novelty | Is the design of the product creative? Does it catch the interest of users? |
| (8) | Stimulation | Is it exciting and motivating to use the product? Is it fun to use? |
| (9) | Clarity | Does the user interface of the product look ordered, tidy, and clear? |
| (10) | Quality of Content | Is the information provided by the product always actual and of good quality? |
| (11) | Immersion | Does the user forget time and sink completely into the interaction with the product? |
| (12) | Aesthetics | Does the product look beautiful and appealing? |
| (13) | Identity | Does the product help the user to socialize and to present themselves positively to other people? |
| (14) | Loyalty | Do people stick with the product even if there are alternative products for the same task? |
| (15) | Trust | Do users think that their data is in safe hands and not misused to harm them? |
| (16) | Value | Does the product design look professional and of high quality? |

In relation to the described approaches, UX factors are typically constructed by using empirical methods of item reduction, such as principal component analysis (PCA). For this, items are grouped into factors based on their empirical correlations. As a result, scales may consist of items that represent, at least at first sight, semantically different concepts. Thus, in some cases, it is not directly clear to describe what the semantic meaning behind a scale is. This provides a completely new perspective on the concept of UX. To get a deeper understanding of the concept of UX, it makes sense to analyze the purely semantic similarities of items and to investigate a structuring based on this concept.

Up to now, only two approaches have conducted the semantic textual similarity in the field of UX research [47][48]. One of the studies is accepted for publication in 2024 [48]. Both studies applied NLP techniques at the level of the measurement items, analyzing the semantic textual similarity between the items. The main goal of both approaches was to conduct a common ground based on semantically similar measurement items. For this, a Sentence Transformer Model and a Sentence Transformer-based Topic Modeling technique were applied to analyze the semantic structure of the textual items [47][48].

The first study by [47] measured the sentence similarity by applying the Sentence Transformer Model Augmented SBERT (AugSBERT), which is a cross- and bi-encoder Transformer architecture [24]. The AugSBERT encodes the textual UX measurement items into embedding in a vector space. Based on the spatial distance, the cosine similarity between the items was calculated, and items were clustered based on a determined similarity threshold. This results in different clusters with semantically similar items [47]. The second study (which is to be published) extends the first procedure by applying the Sentence Transformer-based Topic Modeling BERTopic developed by [49]. The procedure is similar to the first approach, encoding the textual items into embeddings using the SBERT [23]. Based on this, BERTopic clustered the different embeddings [48]. The results of both studies indicate that innovative NLP techniques can be useful in determining semantic textual similarity. However, several weaknesses in both approaches can be recorded. For further insights, we refer to the respective articles [47][48].

Since the release of ChatGPT in November 2022, the development and popularity of GenAI have increased rapidly in various fields, e.g., NLP is revolutionized [50][51]. GenAI can be applied to different tasks ranging from process support to automation to enhance productivity. This article presents an extended approach based on [1] applying GenAI in UX research. For this, we aim to find out whether GenAI can be usefully applied in this field. We used ChatGPT-4 as LLM [52] to (1) (re-)construct UX factors, (2) detect and assign similar items to existing UX concepts, and (3) to analyze the semantic textual similarity of measurement items as well as assign them to the respective similar UX quality aspect. The detailed approach is explained in the following Section IV.

IV. METHODOLOGICAL APPROACH

In this study, we applied a large language model (LLM), which is becoming increasingly popular in both academia and industry. LLMs are statistical language models referring to the following characteristics [53]:

- large-scale
- pre-trained
- transformer-based neural networks

Due to their structure, LLMs indicate strong language understanding and generation abilities. Therefore, complex language tasks can be solved. Moreover, LLMs can be augmented by external knowledge and tools. Thus, LLMs are useful for a broad range of deep learning and natural language processing tasks. This also represents the largest area of application of LLMs in research, as the initial objective in the development of LLMs was to increase the performance of NLP tasks [53]–[57].

Within this domain, LLMs can effectively be used for tasks related to natural language understanding, such as text classification or semantic understanding, referring to the comprehension and interpretation of language based on the underlying semantic meaning and intent (See Section II-B). Previous research indicates the good performance of LLMs regarding these tasks [53][57]. Concerning text classification, Yang and Menczer showed that ChatGPT produced acceptable results in text classification [58]. Even though the capabilities of semantic understanding by LLMs are constrained, they also indicate reasonable results [57].

In this study, we applied an LLM for text classification and semantic understanding, with the main focus on the latter. In particular, ChatGPT-4 was used to analyze UX measurement items to determine similarity topics based on semantically similar items. ChatGPT-4 is a large multimodal model developed by OpenAI. The LLM is based on the Generative Pretrained Transformer architecture GPT-4. For detailed insights, we refer to OpenAI (https://openai.com/gpt-4) [52].

The methodological approach is a four-step procedure consisting of data collection and three investigations using ChatGPT-4. The three investigations consist of text-processing tasks referring to text classification and semantic understanding based on input data and prompting. The detailed approach is described below.

As a first step, data was collected. A set of 40 established UX questionnaires [9] was analyzed. We excluded all questionnaires with (1) a semantic differential scale and (2) a divergent measurement concept, i.e., specifically formulated items focusing on a concrete evaluation objective. This resulted in a list of 19 questionnaires with 408 measurement items. Figure 1 illustrates the data collection process.

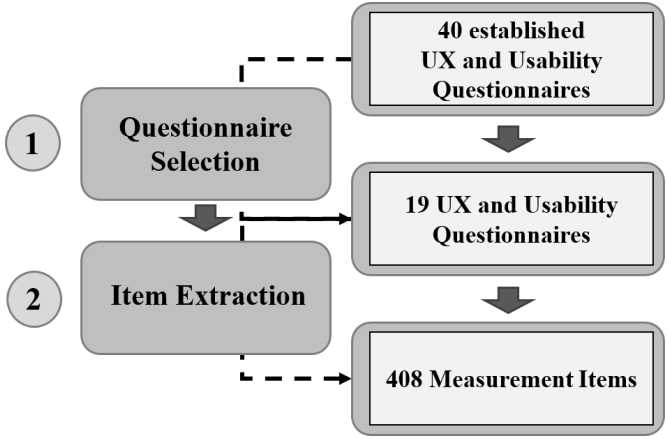

Figure 1: Data Collection.

In the second step, we aimed to gather insights into whether GenAI can perform a (Re-)Construction of UX Factors (see Section V). We introduced all items to ChatGPT-4, and six prompts were formulated. The prompts described the task for the LLM generating topics based on semantically similar items.

In the third step, we aimed to detect suitable items fitting a pre-defined UX concept very well based on the analyzed data set of items from the first step. Therefore, we formulated a generic prompt and adjusted it to each quality aspect to detect appropriate items for existing UX quality aspects (see Section VI). Such detecting and assignment is particularly useful for "ad-hoc surveys" that do not use a standardized questionnaire to measure UX, but just a bunch of self-made questions to find out something specific. This often requires spontaneous additional questions. Thus, before formulating new items, the search and detection of measurement items within an existing item pool using GenAI is quite practical.

In the fourth step, we want to go beyond such detection by analyzing the semantic textual similarity of the UX measurement items. We applied ChatGPT to standardize all items artificially. Afterward, all adjectives of positively formulated items were extracted. Based on this, we again used ChatGPT to analyze the semantic textual similarity of all adjectives. Moreover, semantically similar items were assigned to the respective semantically suitable UX quality aspect. The four-step procedure is visualized in the following Figure 2.

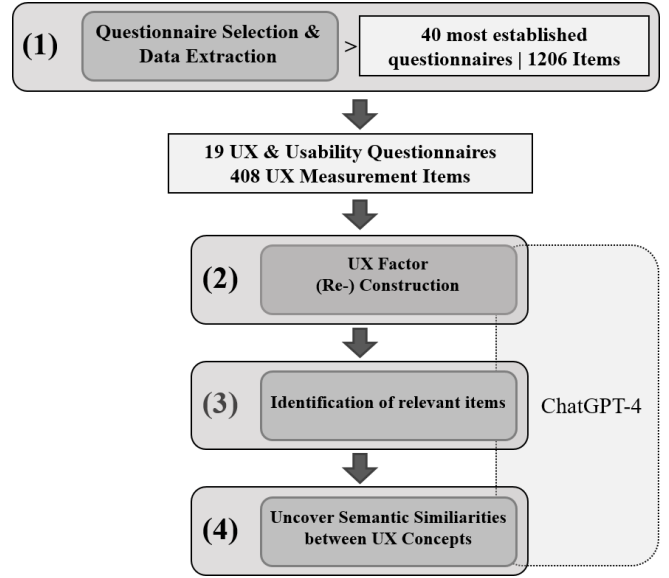

Figure 2: Methodological Approach.

The item detection for all quality aspects concerning the third step as well as step four represent the extension of this approach in relation to [1]. Further details on the procedure and the results of the respective steps are shown in the following Sections V, VI, and VII.

## V. UX Factor (Re-) Construction

### A. Definition of Prompts

After data collection, the second step of the procedure was performed. This first experimental part aims to answer RQ1 whether GenAI can be used to (re-)construct common UX factors. Therefore, ChatGPT was applied to (re-) construct UX factors based on the UX measurement items in relation to their semantic textual similarity. We formulated six prompts. The different tasks given to ChatGPT are described in detail below. The prompts are shown in the following:

- **prompt1**: "Can you extract the questions with a high similarity, i.e., answering about similar topics?"
- **prompt2**: "Can you break this down more detailed?"
- **prompt3**: "Can you try to break down each section into more subsections with its own category?"
- **prompt4**: "Can you improve your categorization?"
- **prompt5**: "In literature, I can find such a list with 16 UX factors.—*inserted the defined quality aspects (see Table I)*—. Can you compare this list with your categorization and contrast these lists?"
- **prompt6**: "I would like you to take your categorization you have done earlier and improve this into more generalized, holistic topics"

At first, a simple classification was performed (*prompt1*). We further tried to break this classification down to determine more specific topics (*prompt2*). In the third step, the topics were divided into subcategories by inserting *prompt3*. *Prompt4* specifies the task of a topic improvement. In particular, the LLM shall optimize the respective topics and subtopics classified so far and, thus, create a further advanced classification. With *prompt5*, we introduced existing UX quality aspects (see Section III) to ChatGPT, comparing them with the AI-generated topics in relation to their similarities and differences. By taking these into account, we lastly aimed to generate and improve the categorizations into more general topics, providing a holistic perspective with *prompt6*. Thus, the formulated prompts mainly refer to exploratory structuring and improvement of the data.

In the following, the different prompts given to ChatGPT and the respective results are presented.

*B. Results*

*1) Prompt1: Primary Classification:* Regarding the first prompt, ChatGPT provided a first classification by themes resulting in six topics. In addition, the respective classified items were assigned to each generated topic. We have only provided the first three most representative items for each category (see Appendix A1). The classification is shown in the following:

- (1) **Usability and Ease of Use**
- (2) **Design and Aesthetics**
- (3) **User Engagement and Experience**
- (4) **Trust and Reliability**
- (5) **Information Access and Clarity**
- (6) **Issues and Errors**

Results show that common topics emerge. Topics with both functional and emotional properties were generated. In relation to the classified items, the generated topics based on the item classification are considered plausible. However, the item formulations are very specific compared to the rather broad generated categorizations. As an example, we can show Topic (1) named **Usability and Ease of Use**. The first three representative items of this topic, however, refer specifically to Ease of Use. Thus, the AI-generated topics from the first step are very broad.

As a result, we can show that ChatGPT can identify logical topics based on the semantic textual structure. However, a classification of six topics based on a total of 408 items is very superficial.

*2) Prompt2: More Detailed Classification:* We proceeded by asking the LLM for a more specific classification, deriving a more detailed classification. Therefore, *prompt2* was applied. As a result, ChatGPT classified ten topics. The respective items of the ten topics can be seen in the Appendix (see A2).

- (1) **Ease of Use**
- (2) **Complexity and Usability Issues**
- (3) **Design and Appearance**
- (4) **Engagement and Immersion**
- (5) **Performance and Responsiveness**
- (6) **Reliability and Trust**
- (7) **Information Quality and Access**
- (8) **Errors and Bugs**
- (9) **Learning and Memorability**
- (10) **Effectiveness and Efficiency**

Referring to the results of the second classification, four more topics are contained and, thus, it is more precious. In more detail, Topic (1) was further divided into two topics compared to *prompt1*. In addition, classifications of Performance and Responsiveness, Learning and Memorability, and Effectiveness and Efficiency were added. Compared to the first classification, the functional, task-related topics were further broken down. Thus, the majority of AI-generated topics relate to a rather pragmatic quality. Topic (1), (2), (5), (7), (8), (9), and (10) are of pragmatic property whereas (3) and (4) are of hedonic property addressing the emotional perception of the user. Moreover, Topic (6) – Reliability and Trust – contains both pragmatic and hedonic items. This indicates that, in general, the measurement items seem to be more pragmatic-oriented among the topics. To conclude, it was possible to distinguish the topics more precisely and categorize them in a more detailed way using the LLM.

However, the classified items within the different topics are still broad concerning the formulation. Some items can be applied to many different scenarios, e.g., *"it meets my needs"*, whereas other items show a high specification, e.g., *"I feel comfortable purchasing from the website"*. Thus, it seems logical to provide an even more detailed categorization into subcategories.

*3) Prompt3: Extended Classification:* By inserting *prompt3* we aimed to generate a more detailed classification within the different topics. We asked ChatGPT for a specific breakdown into subsections resulting in 22 further subtopics:

- **Ease of Use**
  System Usability—Website Usability—Application Usability
- **Complexity and Usability Issues**
  System Complexity—Frustration and Difficulty—System Limitations
- **Design and Appearance**
  Visual Attraction—Layout and Structure—Design Consistency
- **Engagement and Immersion**
  Time Perception and Involvement—Depth of Experience
- **Performance and Responsiveness**
  Speed of Response
- **Reliability and Trust**
  Website Trustworthiness—System Reliability
- **Information Quality and Access**
  Quality of Information—Accessibility of Information
- **Errors and Bugs**
  Technical Issues—Error Messages
- **Learning and Memorability**
  Learning Curve—Recall and Retention
- **Effectiveness and Efficiency**
  Functional Efficiency—Expected Functionality

The generated division into main- and sub-topics confirms the specification and characteristics of the measurement items on different levels. This can be traced back to the different characteristics and focus of the UX questionnaires and their items. All three prompts determined the level of categorization. A further step was taken prompting ChatGPT to make improvements.

*4) Prompt4: Classification Improvement:* Regarding *prompt4*, ChatGPT was given the task of improving the classification without any further specifications. This results in six main topics with 16 subtopics. The number of main topics was reduced. This returns to a rather broad generation of topics. Moreover, a broad spectrum of sub-topics was generated. Concerning the sub-topics, ChatGPT changed the categorizations and classified both pragmatic and hedonic topics together. For instance, **Aesthetics and Design** is grouped with **Navigation and Usability**.

Besides this, the LLM mainly generates suitable topics and respective sub-topics. For instance, the main topic **System Usability and Performance** contains the three sub-topics **Ease of Use, Efficiency and Speed, and Functionality and Flexibility** being purely pragmatic. By comparing this topic generation to the definition by the DIN ISO [2], it mainly captures the whole concept of usability. However, more topics are of pragmatic property than hedonic property.

- **System Usability and Performance**
  Ease of Use—Efficiency and Speed—Functionality and Flexibility
- **User Engagement and Experience**
  Engagement Level—Aesthetics and Design—Confusion and Difficulty
- **Information and Content**
  Clarity and Understandability—Relevance and Utility—Consistency and Integration
- **Website-specific Feedback**
  Navigation and Usability—Trust and Security—Aesthetics and Design
- **Learning and Adaptability**
  Learning Curve—Adaptability
- **Overall Satisfaction and Recommendation**
  Satisfaction—Recommendation

Considering the results, a two-level structure by main and sub-topics is presented. It must be mentioned that some main topics, being rather broad, contain sub-topics with pragmatic as well as hedonic properties. To sum up, ChatGPT generates a useful improvement of topics in general.

*5) Prompt5: Comparison Towards Existing Consolidation:* As we have already described in Section III, some approaches were conducted to consolidate UX factors and find common ground by analyzing semantic and empirical similarity. However, only the former records by [6][13][46] focusing on empirical similarity provided a systematic list of UX factors/UX quality aspects. Thus, a comparison between approaches based on empirical similarity and consolidation based on semantic similarity is useful. For this, we consulted the latest existing UX concepts (see Table I) developed by [13] and compared them to the AI-generated categories. We aimed to compare existing consolidations based on empirical similarities and the topics based on semantic similarities generated by LLM. In particular, we inserted the existing UX quality aspects and formulated the prompt as follows: *"In literature, I can find such a list with 16 UX factors.—inserted the defined quality aspects (see Table I) [13]—. Can you compare this list with your categorization and contrast these lists?"*. The comparison is illustrated in Table II.

TABLE II: COMPARISON OF EXISTING UX QUALITY ASPECTS [13] AND AI-GENERATED TOPICS.

| (#) | UX Quality Aspects | AI-generated Sub-Topics |
|---|---|---|
| (1) | Perspicuity | Ease of Use—Learning Curve |
| (2) | Efficiency | Efficiency and Speed |
| (3) | Dependability | Consistency and Integration |
| (4) | Usefulness | Functionality and Flexibility—Relevance and Utility |
| (5) | Intuitive use | Ease of Use |
| (6) | Adaptability | Adaptability |
| (7) | Novelty | - |
| (8) | Stimulation | Engagement Level |
| (9) | Clarity | Clarity and Understandability |
| (10) | Quality of Content | Relevance and Utility |
| (11) | Immersion | Engagement Level |
| (12) | Aesthetics | Aesthetics and Design—Aesthetics and Design |
| (13) | Identity | - |
| (14) | Loyalty | Loyalty |
| (15) | Trust | Trust and Security |
| (16) | Value | Perceived Value |

Before considering the results, it must be noted that the quality aspects by [13] do not consist of sub-topics. Results show some fundamental differences. Firstly, it must be stated that the LLM did not allocate all AI-generated topics to the existing quality aspects. In particular, the categorization does not include the UX quality aspects of *Novelty* and *Identity*. Furthermore, specific items and factors overlap as some AI-generated factors were allocated to more than one quality aspect. In general, the consolidation by [13] (see Table I) is more generalized without a focus on a specific interactive product. For instance, the LLM categorized the sub-topic *Trust and Security* in the main topic *Website-specific Feedback*. This indicates that *Trust and Security* specifically refers to the context of Websites. In contrast, *Trust* as a stand-alone quality aspect by [13] is defined more generally. To conclude, UX quality aspects based on former approaches concerning empirical similarity indicate a more holistic view covering both pragmatic and hedonic aspects of UX, whereas the AI-generated topics and sub-topics show a stronger focus on the pragmatic property as well as a deeper focus on specific products. Due to a high degree of specification, problems with general applicability may arise. Nevertheless, there are many similarities between the two consolidations, and thus, the AI-generated topics by ChatGPT can be considered logical.

*6) Prompt6: Construction of Generalized Categories:* Based on the former results, there is still a lack of certain generality and focus on hedonic properties within the AI-generated categories. For this, *prompt6* was formulated to create more generalized topics and, thus, to provide a more holistic view of UX. We prompted as follows: *"I would like you to take your categorization you have done earlier and improve this into more generalized, holistic topics"*. In this context, it is important to see which items represent the AI-generated topics, as the consolidation is originally based on the semantic similarity of the measurement items. We prompted

ChatGPT to issue the top five items representing the respective topic best. Concerning the results, the LLM generates a comprehensive overview with generalized UX factors as well as their definitions and items. A two-dimensional separation into the main topic and sub-topics can be shown. Additionally, both pragmatic and hedonic properties are contained. Thus, ChatGPT provides a comprehensive and generalized view of the construct of UX.

In particular, ChatGPT generated six main topics and 15 sub-topics (see Appendix A3). Concerning the results, the consolidated and AI-generated topics concerning a holistic view of UX fit well compared to previous research. Thus, the LLM is useful in deriving general UX concepts based on AI-generated topics. Pragmatic and hedonic properties are captured. The items are almost entirely coherent with each other and fit the construct. In particular, pragmatic topics show high similarities to existing literature and can be considered as well generated. However, applying ChatGPT still faces some weaknesses. For instance, different classifications of items differ quite strongly and are accordingly not representative of the respective topic. In this context, the topic *Identity* can be listed. In addition, items (4) and (5) (see Appendix A3) categorized in **Consistency and Integration** must be mentioned. The item's property is hedonic, whereas the topic and classified items (1)-(3) are considered pragmatic. Thus, a semantic relation between obviously pragmatic and hedonic items can be indicated. This coincides with previous research (see Section III). To illustrate the fit between item property and topic characteristics, we added a **(+)** for a suitable item fit and a **(-)** for an unsuitable item fit. It also may be that some items are contained in multiple topics due to a rather general formulation. In this case, the researchers added **(+-)** (see Appendix A3).

## VI. IDENTIFICATION OF RELEVANT ITEMS

Up to this point, we have demonstrated how GenAI can be used to define a semantic structure on a large set of items from UX questionnaires. Another quite natural use case is to detect those items that best represent a clearly defined UX concept. In this section, we provide several examples to illustrate this.

### A. Definition of a Generic Prompt

We use a special prompt (in the following referred to as *prompt7*) for this purpose. On top of the prompt, there was a short instruction and explanation of a typical UX concept.

For example, for Learnability (how easy or difficult it is to get familiar with a product) the corresponding instruction was:

*"Below there is a list of statements and questions related to the UX of a software system. Select all statements or questions from this list that describe how easy or difficult it is to learn and understand how to use the software system. List these statements or questions. Start with those statements and questions that describe this best.*

The list of 408 items from UX questionnaires was placed directly below this instructional part of the prompt.

This prompt can easily be adapted to represent other UX concepts if the part "Select all statements or questions from this list that describe how easy or difficult it is to learn and understand how to use the software system." is replaced by another formulation.

### B. Results

For this example, the resulting list contained items that refer to ease of learning *(It was easy to learn to use this system)*, intuitive understanding *(The system was easy to use from the start)*, or aspects that support the user to handle the product *(Whenever I made a mistake using the system, I could recover easily and quickly)*.

The top 10 items filtered out for Learnability are:

1) It was easy to learn to use this system
2) I could effectively complete the tasks and scenarios using this system
3) I was able to complete the tasks and scenarios quickly using this system
4) I felt comfortable using this system
5) The system gave error messages that clearly told me how to fix problems
6) Whenever I made a mistake using the system, I could recover easily and quickly
7) The information provided with this system (online help, documentation) was clear
8) It was easy to find the information I needed
9) The information provided for the system was easy to understand
10) The information was effective in helping me complete the tasks and scenarios

Thus, the detected items fit well with the request in the prompt.

To assess the quality of other UX concepts, we modified the prompt by using various replacements for the variable part mentioned earlier. We explored the following additional UX concepts:

- **Efficiency:** Select all statements or questions from this list that describe how efficient or inefficient it is to work with the software system.
- **Usefulness:** Select all statements or questions from this list that describe whether the software system is useful or not.
- **Dependability:** Select all statements or questions from this list that describe if the user feels in control when he or she works with the software system or if this is not the case.
- **Stimulation:** Select all statements or questions from this list that describe how stimulating or boring it is to work with the software system.

Appendix A4 shows the top 10 items per concept. Again, the detected items fit well with the UX concepts described in the prompt.

However, there are some differences that must be highlighted. For the classical UX concepts of Efficiency, Usefulness, and Dependability, the top 10 items showed a strong alignment with these concepts. There are a few exceptions that would be classified differently by a UX expert. For example, *The processing times of the software are easy for me to estimate* was classified under Efficiency, but it is a classical item that reflects Dependability (does the user feel in control and can predict the behavior of the system). This misclassification may be due to the presence of the words *processing times*. Similarly, items 9 and 10, which were assigned to Usefulness, are more closely related to Dependability.

In terms of Stimulation, some of the items were a good fit for the concept, particularly the first four. However, the remaining items did not adequately capture the essence of Stimulation. This can be attributed to the fact that our initial item set was derived from older questionnaires that primarily focused on usability, neglecting hedonic aspects like Stimulation. Therefore, it is not surprising that the language model selected these rare examples, while the rest of the chosen items only loosely corresponded to Stimulation. This example clearly demonstrates that language models can assist UX researchers in identifying suitable items, but it is crucial to evaluate the results and make necessary corrections critically.

## VII. Uncover Semantic Similarities between common UX Concepts

In our first two investigations, we utilized a collection of items derived from traditional usability questionnaires. We had to omit semantic differentials due to their distinct format compared to the statement-based items, which poses challenges for automatic analysis by a language model. For our third study, we created a new item set.

The items have been artificially created in order to achieve a highly standardized format, which would not have been possible if we had directly selected them from UX questionnaires. Each item follows the structure "I perceive the product as <adjective>". For example, "I perceive the product as efficient" or "I perceive the product as exciting". Only positive adjectives are used. The adjectives were extracted from existing items in UX questionnaires using two methods. For semantic differentials, simply the positive term was taken (for example, from inefficient/efficient, we take the positive term efficient). For items represented as statements, we removed all other parts of the item and kept only the positive adjective. If the item has a negative formulation, i.e., there is no positive adjective, the item is ignored. For example, the item "Is the cursor placement consistent?" is transformed into "I perceive the product as consistent".

In total, 135 artificial items could be constructed. See [8][59] for a similar technique to display typical UX items from standardized questionnaires as a word cloud.

### A. Definition of a Generic Prompt

We use a standard prompt (referred to in the following as *prompt8*) to filter those items that correspond to a typical UX concept. On top of the prompt, there was a short instruction and explanation of a typical UX concept. For example, for Learnability the corresponding instruction was:

*Below there is a list of statements related to user experience of a product. Select all statements from this list that describe that it is easy to learn and to understand how to use the product. List these statements or questions. Start with those statements and questions that describes this best.*

The list of 135 artificial items was placed directly below this instructional part of the prompt.

For other UX concepts, the part *Select all statements from this list that describe that it is easy to learn and to understand how to use the product* was replaced. The rest of the prompt stays stable.

The following replacements were used:

- **Learnability:** Select all statements from this list that describe that it is easy to understand and to learn how to use the product.
- **Efficiency:** Select all statements from this list that describe that users can solve their tasks using the product efficiently without unnecessary effort and that the product reacts fast on user commands or data entries.
- **Dependability:** Select all statements from this list that describe that the user feels in control of the interaction and think it is secure and predictable.
- **Stimulation:** Select all statements from this list that describe that it is exciting, motivating and fun to use the product?
- **Novelty:** Select all statements from this list that describe that users perceive the product as original and creative.
- **Aesthetics:** Select all statements from this list that describe that the product looks beautiful, aesthetic and appealing.
- **Adaptability:** Select all statements from this list that describe that the user perceives that the product can be easily adapted to his or her personal preferences or working styles.
- **Usefulness:** Select all statements from this list that describe that users perceive the product as useful.
- **Value:** Select all statements from this list that describe that the product design looks professional and of high quality.
- **Trust:** Select all statements from this list that describe that the users think that their data are in safe hands and are not misused.
- **Clarity:** Select all statements from this list that describe that users think that the user interface of the product looks ordered, structured, and is of low visual complexity.

Each prompt was utilized in three separate runs of ChatGPT-4. For the final analysis, we only considered items that were consistently assigned to the concept in all three runs.

### B. Results

The following graphic depicts the results (see Figure 3). The words in upper case font represent the UX concepts. Lowercase font the adjectives of the items (rest removed to avoid clutter). A line shows if an adjective was related to a UX concept by ChatGPT.

Figure 3: Uncover Semantic Similarities.

On the left side of the chart, we see the hedonic UX aspects Novelty, Stimulation, Aesthetics, and Value. Stimulation and Novelty do not share any item with other UX concepts, i.e., they represent semantically clearly distinct properties. Aesthetics and Value share a lot of items, they have a huge semantic overlap. This is a quite natural result. Value represents the feeling that a product looks professional and of high quality. But of course, a product that does look aesthetically unappealing will not be regarded as professional or of high quality. Trust is more or less isolated, but shares one item with Dependability. Adaptability is connected to Dependability and Efficiency. Usefulness is heavily connected to Efficiency. Efficiency and Learnability are connected by just one item, while both are heavily connected to Dependability. A very interesting observation is the indirect connection between Aesthetics and the classical usability criteria of Efficiency, Dependability, and Learnability. This connection is established over Value and Clarity. This fits well with empirical studies [27] that showed that Clarity is a mediator variable that explains the dependency between Aesthetics and classical usability dimensions.

Of course, we should be careful not to over-interpret these results. The outcome might be different if we modify the formulations in the prompts and of course, also depend on the version of the used LLM. However, such analyses are quite useful for understanding what typical UX concepts mean semantically and how much they overlap.

Another interesting question is how well the selected items fit empirically constructed scales. Most of the UX aspects used in this investigation correspond to scales in the UEQ or UEQ+. For the scale construction in those questionnaires, pools of items were created, data were collected from participants that evaluated different products with all items from the item pool, a principal component analysis was performed, and the four best-fitting items per component were then selected to represent the scale [39][44]. Not all adjectives used in our semantic analysis were contained in these item pools and the same is true vice versa. Thus, we can not expect a perfect match, but it is worth checking how close the empirically constructed scales are to the semantic analysis.

We list these scales and the positive term from the corresponding items (semantic differentials) in the following. The term is bold if it is also assigned to the corresponding category in our semantic analysis.

- Efficiency: **fast**, **efficient**, **practical**, organized
- Learnability (Perspicuity): **understandable**, **easy to learn**, **clear**, **easy**
- Dependability: **predictable**, supportive, secure, meets expectations
- Stimulation: valuable, **exciting**, **interesting**, **motivating**
- Novelty: **creative**, inventive, leading edge, **innovative**
- Aesthetics: **beautiful**, **stylish**, **appealing**, **pleasant**
- Adaptability: adjustable, changeable, **flexible**, extendable
- Usefulness: **useful**, **helpful**, **beneficial**, rewarding
- Value: valuable, presentable, **tasteful**, **elegant**
- Clarity: well-grouped, **structured**, ordered, **organized**
- Trust: **secure**, **trustworthy**, **reliable**, transparent

The correspondence between the semantically constructed item assignment and the empirical assignment is remarkably close for most categories. Even in cases where the items do not fully match, a comparison reveals a high degree of similarity. However, there are a few rare exceptions. For instance, the term *valuable* is represented in the UEQ+ as part of the UX scale Value (which is not surprising). In the conducted semantic analysis, it is assigned to Usefulness, which is somewhat less natural. While the overall fit to empirically constructed scales is good, there are a few exceptions that would benefit from careful review by a human expert to improve the results.

## VIII. Conclusion and Future Work

In this research, we present a GenAI-based approach concerning UX research. The article aims to investigate the usefulness of GenAI in this research field. We applied the LLM ChatGPT-4 to analyze two pools of UX items from established UX questionnaires concerning three different approaches. In particular, we conducted whether GenAI can (1) (re-) construct common UX factors, (2) detect similar items, and (3) cover the semantic similarity as well as assign adjectives to semantic similar UX concepts.

### A. Implications

We showed that LLMs can be usefully applied in UX research. ChatGPT was able to (1) (re-) construct and classify UX factors, (2) detect and assign similar items to the respective quality aspects, and (3) identify the semantic textual structure of the measurement items as well as assign semantic similar items to the suitable quality aspects. To conclude, applying ChatGPT was useful for conducting all three tasks. The three research questions (see Section I) can be confirmed. Thus, applying GenAI in the field of UX enhances research.

However, LLMs are inherently non-deterministic models. Hence, applying the same sequence of prompts once again, the resulting classifications will differ. Nevertheless, this is no problem as there is no objectively "correct" classification of UX factors. Compared to the practice, conducting the same task independently by several UX experts will also result in different classifications. By applying GenAI for this task, however, the effort required for such an automatic classification is extremely low. Thus, the possibility to create such classifications quickly and efficiently allows an explorative search for semantic structures in large sets of items, uncovering interesting hidden dependencies that would be hard to detect with a manual analysis by UX experts.

Considering the results regarding the UX factor (re-) construction, ChatGPT generated a consolidated list of topics, subtopics, and items representing the concept of UX comprehensively. Within the AI-constructed topics, both pragmatic and hedonic aspects were contained. By comparing AI-generated topics with existing UX concepts, a good alignment can be illustrated. In relation to the second task, semantically similar items were detected and assigned to the existing quality aspects based on their respective definition. Regarding the third task, the LLM was useful in uncovering the semantic textual similarity of the items and assigning them to the respective UX concept.

## B. Limitations and Future Research

Concerning this approach of this paper, several limitations must be drawn. Within the first step of data collection (see Section IV), semantic differentials that are a quite common item format in UX questionnaires must be excluded from the analysis to ensure at least a low level of item comparability. This mainly concerns the steps of UX factor (re-) construction and item identification. By including all formats of items, the LLM may achieve even better results.

Future research in prompt engineering shall investigate the possibility of allowing a combination of all common item formats in one analysis. Moreover, analyzing the semantic textual similarity and comparing common UX concepts (see Section VII) provides the possibility of breaking down the construct of UX in a new way.

From a practical perspective, GenAI can be usefully applied for different tasks in UX evaluation scenarios in general. More specifically, the different UX evaluation methods and their respective procedure steps must be analyzed. Based on this, the context and tasks in which GenAI is practicable and applicable must be identified. Afterward, the application within the various scenarios must be tested.

The results of this approach can be taken as a measurement framework for quantitative UX evaluation. Moreover, a UX questionnaire can be derived from the AI-generated topics and the respective items in relation to semantic textual similarity. This results in the first AI-generated UX questionnaire, which is also the first constructed UX questionnaire based on semantic similarity instead of empirical similarity. Furthermore, a comprehensive item list could be detected so that researchers do not have to develop new items but can instead use the existing pool. Thus, providing suitable measurement items quickly and easily would enhance UX evaluation and help researchers. At least, the AI-generated items could be further validated to compromise valid, reliable, and useful results.

This approach is a further step towards a common ground in UX research on the level of the measurement items. The fundamental difference between empirical and semantic similarity is to be emphasized. Moreover, this work can be seen as a first step towards a new research agenda in the field of UX.

## APPENDIX

**A1: Respective first three allocated items of AI-generated topics prompt1:**

**Usability and Ease of Use**
The system is easy to use.
I found the system unnecessarily complex.
I thought the system was easy to use.

**Design and Aesthetics**
The design is uninteresting.
The design appears uninspired.
The color composition is attractive.

**User Engagement and Experience**
I felt calm using the system.
I was so involved in this experience that I lost track of time.
I lost myself in this experience.

**Trust and Reliability**
I feel comfortable purchasing from the website.
I feel confident conducting business on the website.
It is a site that feels secure.

**Information Access and Clarity**
I am able to get the information I need easily.
Provides quick and easy access to finding information.
Provides relevant information.

**Issues and Errors**
The system is too inflexible.
The interaction with the system is irritating.
The interaction with the system is frustrating.

**A2: Respective first three allocated items of AI-generated topics of prompt2:**

**Ease of Use**
The system is easy to use.
I thought the system was easy to use.
This system is easy to use.

**Complexity and Usability Issues**
I found the system unnecessarily complex.
I think that I would need the support of a technical person to be able to use this system.
I found this application confusing to use.

**Design and Appearance**
The design is uninteresting.
The design appears uninspired.
The color composition is attractive.

**Engagement and Immersion**
I felt calm using the system.
I was so involved in this experience that I lost track of time.
I lost myself in this experience.

**Performance and Responsiveness**
The software responds too slowly.
This software responds too slowly to inputs.
The interaction with the system is fast.

**Reliability and Trust**
I feel comfortable purchasing from the website.
I feel confident conducting business on the website.
It is a site that feels secure.

**Information Quality and Access**
I am able to get the information I need easily.
Provides quick and easy access to finding information.
Provides relevant information.

**Errors and Bugs**
The system is too inflexible.

The interaction with the system is irritating.
The software has at some time stopped unexpectedly.

**Learning and Memorability**
I think that I would need the support of a technical person to be able to use this system.
I easily remember how to use it.
It is easy to learn to use it.

**Effectiveness and Efficiency**
It helps me be more effective.
It meets my needs.
It does everything I would expect it to do.

**A3: Generalized AI-generated UX topics with their respective top-5 items:**

**Usability and Intuitiveness**

Ease of Use:
1) The system is easy to use. (+)
2) I think that I would need the support of a technical person to be able to use this system.(+)
3) I would imagine that most people would learn to use this system very quickly.(+)
4) I learned to use it quickly.(+)
5) I can use it without written instructions.(+)

Efficiency and Speed:
1) The interaction with the system is fast.(+)
2) The system responds too slowly.(+)
3) This software responds too slowly to inputs.(+)
4) The speed of this software is fast enough.(+)
5) Has fast navigation to pages.(+)

Adaptability:
1) The system is too inflexible.(+)
2) This software seems to disrupt the way I normally like to arrange my work.(+)
3) It is flexible.(+)
4) It requires the fewest steps possible to accomplish what I want to do with it.(+- Efficiency)
5) It is relatively easy to move from one part of a task to another.(+- Efficiency)

**Content Quality and Clarity**

Relevance and Utility:
1) Provides relevant information.(+)
2) It meets my needs.(+)
3) It is useful.(+)
4) Provides information content that is easy to read.(+)
5) It does everything I would expect it to do.(+)

Consistency and Integration:
1) I thought there was too much inconsistency in this system.(+)
2) I found the various functions in this system were well integrated.(+)
3) I don't notice any inconsistencies as I use it.(+)
4) Everything goes together on this site.(+-)
5) The site appears patchy.(+-)

Clarity and Understandability:
1) The way that system information is presented is clear and understandable.(+)
2) Provides information content that is easy to understand.(+)
3) I think the image is difficult to understand.(+)
4) The layout is easy to grasp.(+)
5) I do not find this image useful.(-)

**Engagement and Experience**

Engagement Level:
1) I was so involved in this experience that I lost track of time.(+)
2) I lost myself in this experience.(+)
3) I was really drawn into this experience.(+)
4) I felt involved in this experience.(+)
5) I was absorbed in this experience.(+)

Stimulation:
1) This experience was fun.(+)
2) I continued to use thr application out of curiosity.(+)
3) Working with this software is mentally stimulating.(+)
4) I felt involved in this experience.(+)
5) During this experience I let myself go.(+- Engagement Level)

Aesthetics and Design:
1) This application was aesthetically appealing.(+)
2) The screen layout of the application was visually pleasing.(+)
3) The design is uninteresting.(+)
4) The layout appears professionally designed.(+)
5) The design appears uninspired.(+)

**Trust and Reliability**

Trust and Security:
1) I feel comfortable purchasing from the website.(+)
2) I feel confident conducting business on the website.(+)
3) Is a site that feels secure.(+)
4) Makes it easy to contact the organization.(+)
5) The website is easy to use.(-)

Dependability:
1) This software hasn't always done what I was expecting.(+)
2) The software has helped me overcome any problems I have had in using it.(+)
3) I can recover from mistakes quickly and easily.(+)
4) I can use it successfully every time.(+)
5) Error messages are not adequate.(+)

**Novelty and Identity**

Novelty:
1) The layout is inventive.(+)

2) The layout appears dynamic.(-)
3) The layout appears too dense.(-)
4) The layout is pleasantly varied.(-)
5) The design of the site lacks a concept.(-)

Identity:
1) Conveys a sense of community.(+)
2) The offer has a clearly recognizable structure.(-)
3) Keeps the user's attention.(-)
4) The layout is not up-to-date.(-)
5) The design of the site lacks a concept.(-)

**Value and Loyalty**

Perceived Value:
1) I consider my experience a success.(+)
2) My experience was rewarding.(+)
3) The layout appears professionally designed.(+)
4) The color composition is attractive.(+)
5) It is wonderful.(+)

Loyalty:
1) I would recommend the application to my family and friends.(+)
2) I would recommend this software to my colleagues.(+)
3) I will likely return to the website in the future.(+)
4) I think that I would like to use this system frequently.(+)
5) I would not want to use this image.(+)

**A4: Top 10 items filtered for additional UX concepts**

**Efficiency**
1) When I work on tasks with the software, I often need more time than planned.
2) I sometimes have to search for a long time for functions that I need for my work.
3) Working with this software is sometimes cumbersome.
4) The software forces me to perform superfluous steps.
5) There are too many input steps to complete some tasks.
6) The system can only be operated in a rigidly predefined manner.
7) The processing times of the software are easy for me to estimate.
8) The software makes my task processing difficult due to inconsistent design.
9) System errors (e.g., "crash") occur during my work with the software.
10) In an error situation, the software provides concrete information on how to correct the error.

**Usefulness**
1) The software helps me to complete my work task better than expected without extra effort.
2) With the software, I can sometimes even exceed my desired goals without any extra effort.
3) The software allows me to increase the quality of my work without any extra work.
4) The software offers me all the possibilities I need to work on my tasks.
5) The software is tailored to the tasks I need to work on.
6) The software allows me to enter data as required by the task.
7) The software offers me a repeat function for recurring work steps.
8) Even non-routine work tasks can be easily processed with the software.
9) The software provides me with information about the current operation and usage options on request.
10) The software provides sufficient information for me about which inputs are currently permitted.

**Dependability**
1) I felt in control of the interaction with the system.
2) The system didn't always do what I wanted.
3) The system didn't always do what I expected.
4) The interaction with the system is unpredictable.
5) The system can only be operated in a rigidly predefined manner.
6) The software forces me to perform superfluous steps.
7) The software allows me to interrupt the editing step, although it expects an input.
8) It is possible to abort at any time when entering a command.
9) The software offers me the possibility to jump from any menu level directly back to the main menu.
10) The software offers me the possibility of customization (e.g., in menus, screen displays) to my individual needs and requirements.

**Stimulation**
1) I sometimes forget the time when I work with the software.
2) The software also allows me to approach my tasks creatively.
3) When I have some free time, I just play around with the software.
4) Even if my actual task is already done satisfactorily, I sometimes try to make it even better with the help of the software.
5) The software forces me to perform superfluous steps.
6) Working with this software is sometimes cumbersome.
7) The system can only be operated in a rigidly predefined manner.
8) The software makes my task processing difficult due to inconsistent design.
9) The product exhilarates me.
10) The product relaxes me.